# DOMAIN-PARTITIONED ELEMENT MANAGEMENT SYSTEMS EMPLOYING MOBILE AGENTS FOR DISTRIBUTED NETWORK MANAGEMENT

Anish Saini[1] & Atul Mishra[2]

[1]Assistant Professor, Department of Computer Science & Engineering, Echelon Institute of Technology, Faridabad, INDIA
[2]Associate Professor, Department of Computer Engineering, YMCA University of Science & Technology Faridabad, INDIA

*ABSTRACT*

*Network management systems based on mobile agents are efficiently a better alternative than typical client/server based architectures. Centralized management models like SNMP or CMIP based management models suffer from scalability and flexibility issues which are addressed to great extent by flat bed or static mid-level manager models based on mobile agents, yet the use of mobile agents to distribute and delegate management tasks for above stated agent-based management frameworks like initial flat bed models and static mid-level managers cannot efficiently meet the demands of current networks which are growing in size and complexity. In view of the above mentioned limitations, we proposed a domain partitioned network management model based-on mobile agent & Element Management Systems in order to minimize management data flow to a centralized server. Intelligent agent allocated to specific EMS performs local network management and reports the results to the superior manager and finally the global manager performs global network management using those submitted management results. Experimental results of various scenarios of the proposed model have been presented to support the arguments given in favor of the prototype system based on mobile agents.*

## 1. INTRODUCTION

The need for data communication has evolved rapidly since the earliest days of computing. Industrial enterprises are increasingly dependent upon networked system serving their information backbones. A typical organization model of a network management system is based on SNMP[1][2] Client/Server architecture. It consists of two major components: network agent process and the network manager process. The network agent process resides on the managed network devices such as routers, switches, servers etc. The network manager is housed on the NMS station from where it manages the various devices, by accessing the management information, through the agents residing on them as shown in Figure 1. The management information consists of collection of managed objects, stored in Management Information Base (MIB). Based on SNMP model, the management of networks from the Network Operation Centers (NOC) is mostly done by a Network Manager which deploys several element management system(EMS). EMSs in turn manage their specific zones or domains. These EMSs provide ISO's five functional categories: Fault Management, Configuration Management, Accounting, Performance Management and Security Management. Network Manager acts as a client to the EMSs, which in turn act as network manager to network devices for retrieval and provisioning of data. The data retrieved by EMSs are stored in databases on the EMSs platforms and used for management of network devices

These existing management models traditionally adopt a centralized, Client/Server (C/S) approach wherein the management application, with the help of a manager, acting as clients, periodically accesses the data collected by a set of software modules, agents, placed on network devices by using an appropriate protocol.

These centralized architectures suffer from the lack of scalability and flexibility. Furthermore, the staleness of gathered data (due to network latency involved) and probable error in the selection of management task being carried over (owing to the staleness of data) reduces the reliability of the management applications. Thus, controlling and managing the traffic in these networks is a challenging task [3].

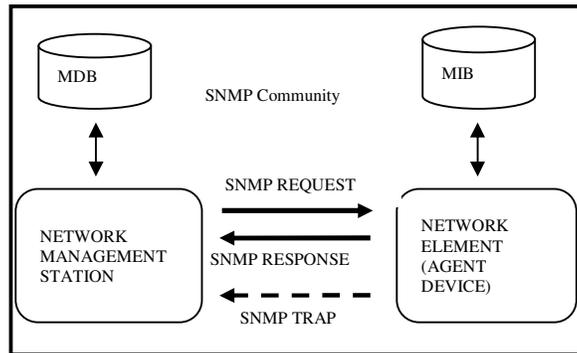

Figure 1. SNMP Model

In view of the above mentioned limitations, we proposed a domain partitioned network management model based-on mobile agent & Element Management Systems in order to minimize management data flow to a centralized server. Intelligent agent allocated to specific EMS performs local network management and reports the results to the superior manager and finally the global manager performs global network management using those submitted management results.

The Mobile Agent (MA) paradigm has emerged within the distributed computing field. The term MA refers to autonomous programs with the ability to move from host to host to resume or restart their execution and act on behalf of users towards the completion of a given task.[4] One of the most popular topics in MA research community has been distributed NM [5][6][7], wherein MAs have been proposed as a means to balance the burden associated with the processing of management data and decrease the traffic associated with their transfers (data can be filtered at the source).

Goldszmidit et al [8], introduces the concept of management by delegation, the management station can extend the capability of the agents at runtime thereby invoking new services and dynamically extending the ones present in the agent on the device. Mobile agent based strategies have distinct advantages over the others as it allowed for easy programmability of remote nodes by migrating and transferring functionality wherever it is required.

Bellavista[9] et al. proposed a secure and open mobile agent environment, MAMAS (Mobile Agents for the Management of Applications and Systems) for the management of networks, services and systems. Sahai & Morin [10] introduce the concept of mobile network managers (MNM), which is a location independent network manager and assists the administrator to remotely control his/her managed network, through launching MAs to carry out distributed management tasks. In, Oliveira and Lopes propose how the integration of MA-based sub-system could be carried out in the IETF's DISMAN framework. In [11], I. Satoh proposes how a network and application independent MA based framework could be designed. Manoj Kumar Kona et al. [12] described an SNMP based efficient mobile agent network management structure, in order to cooperate with conventional management system;

For transferring less network monitoring data and managing devices more effectively, Damianos Gavalas et al.[5] propose a scalable and flexible MA based platform for network management; MAGENTA (Mobile AGENT environment for distributed Applications) introduces Mobile Network Manager (MNM) which is a location independent network manager [13].

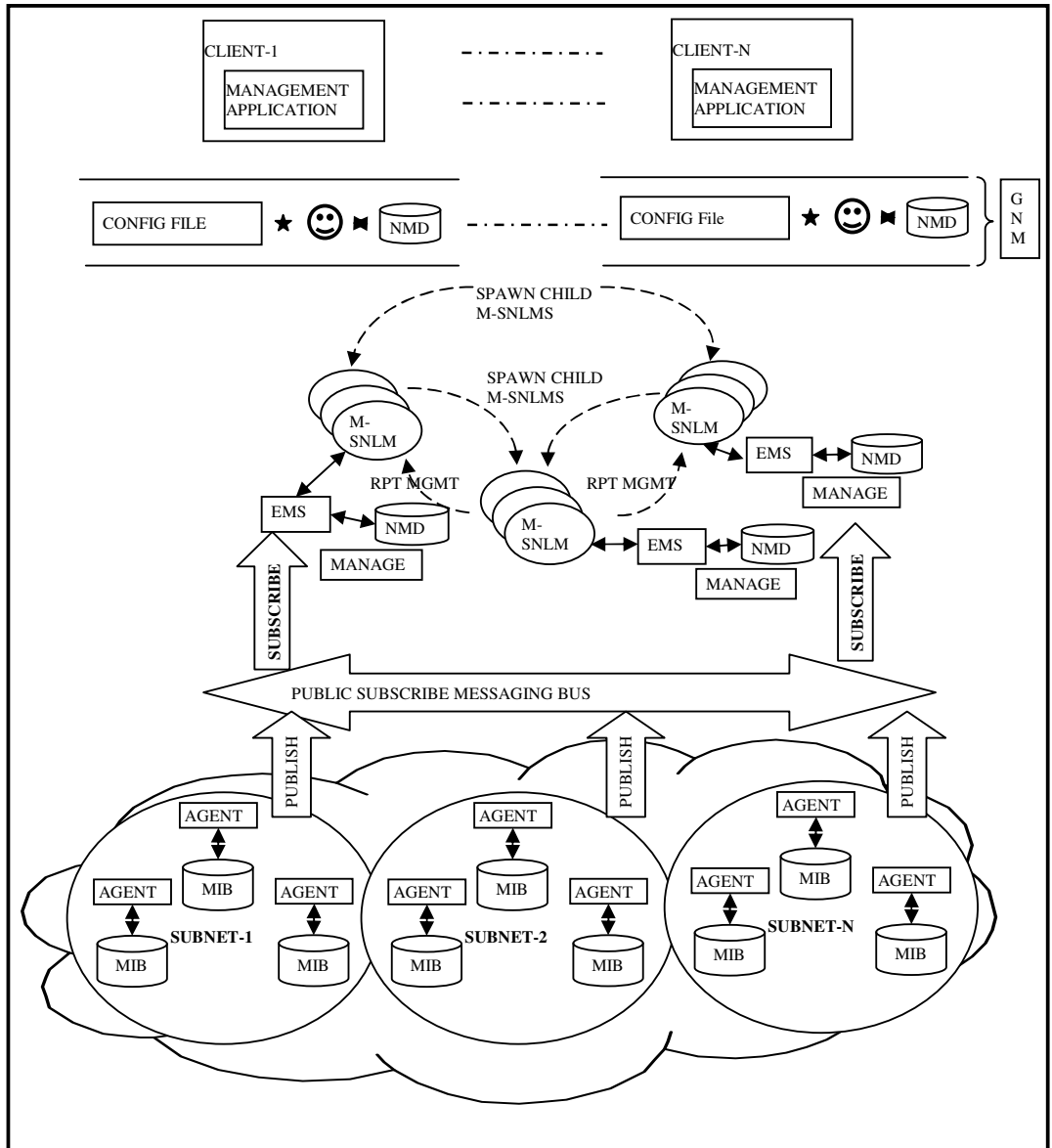

Figure 2.  Management Entity & Interaction Model

The network managers in this architecture utilize client- server technology and/or mobile agent technology as and when required depending on the functionality implemented and their location.

Damianos Gavalas et al.[4] proposed a hierarchical and scalable management model where middle managers are themselves mobile and based on certain policies they dynamically segment the network and deploy other mobile middle managers for data collection.

## 2. HIGH LEVEL DESIGN OF EMS BASED NETWORK MANAGEMENT MODEL DEPLOYING MOBILE AGENTS

A.K. Sharma et al. proposed IMASNM model [16] which discusses strategies for large scale network partitioning, fixing of management scope and deployment of mobile M-SNLMs in various sub-network domains. In IMASNM hierarchal network management model, the mobile managers (M-SNLMs) move in their domains and manages network device with a flat bed model scheme. The manager works with quite good efficiency if the size of domain falls in a specified range. If the domain size increases from that specified range then efficiency of network manager goes down due to management cost of flat bed model which is the main problem of IMASNM model.

Considering above, we proposed an EMS based mobile agent network management model shown in Figure 2. In this scheme, the managed network is divided into many domains based on the geographical layout of the network or number of nodes a manager can efficiently manage or average load on the entire network or certain kind of administrative relationship. In each domain, an Element manager is appointed and for the overall managed network a set of various management applications along with a network manager acting as a Global Network Manager (GNM) are appointed. Element managers (EM) exist at the lowest level of the managers' hierarchy. Each Element manager controls and monitors a set of network element (SONET/DWDM/Ethernet) via specialized protocol independent mediators as shown in Figure 3. After initial discovery of the network, EMS keeps database changes in the sub-network by means of publish/subscribe paradigm. Additionally a hierarchy of

Mobile-Subnetwork Layer Manager (M-SNLM) is appointed in between the GNM and leaf-level EMSs. First levels of M-SNLMs (appointed by GNM) are dispatched to platforms where EMSs are running. M-SNLMs along with EMSs take over a portion of the network from their parent M-SNLMs and act as local managers for that portion of the network for all management needs. Depending upon the growth in their domain, M-SNLMs spawn additional child M-SNLMs for scalable management of the network. M-SNLMs not only interact with the EMSs they are assigned to but also could visit other EMSs platforms depending upon the need as they are mobile agents in themselves

*Highlights of the Proposed Model*

EMS based domain partitioned network management system based-on mobile agent has many technology advantages over conventional network fault management system and other MA based network management models presented by researchers.

- ❖ Earlier work presented by researchers [5][16] assumed the presence of mobile agent runtime environment on the network devices. This may not be true for many telecomm elements and legacy devices. Proposed model make the architecture independent of this constraint.
- ❖ Earlier models [5][16] dispatch data agents to devices in the sub-network and bear a cost associated with it. Proposed model saves this cost by retrieving that data from the database present on the EMS platform only. It also controls the number of mobile agents present in the system as a given M-SNLM can not only manage its assigned domain but can also visit other domains and interact with EMSs to manage the network.

- ❖ Bottleneck problem of convention centralized network management are addressed and static communication problems in distributed network is solved efficiently. The basic concept which is adopted in this framework is: "M-SNLMs solve problems locally and the supervisors/parent M-SNLMs worry about the issues which are propagated to them."

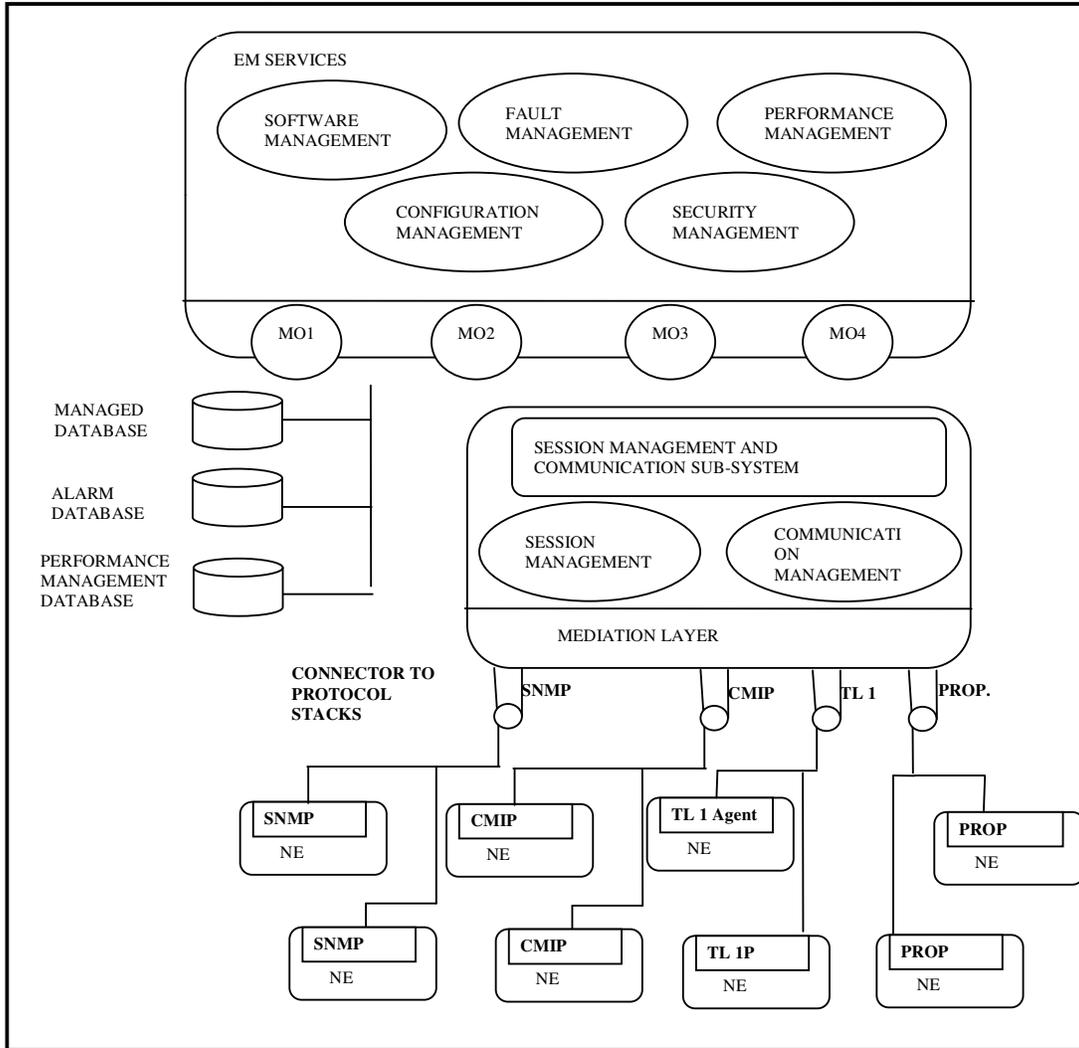

Figure 3. Management Entity & Interaction Model

- ❖ The proposed architecture is based on mobile agent technology with the advantages:

  - ➢ Balancing of Network loads,
  - ➢ The framework is adaptive with network size change,
  - ➢ The feature of mobile agents disconnecting with GNM when working on managed devices saves bandwidth effectively. In addition, mobile agents are endurable with network state and they can execute continually after network recovery, it is also what mobile agent technology exceeds other network management technologies.
- ❖ The proposed architecture is based on mobile agent technology Proposed model offers a hierarchical network management model that can reduce network management complexity and minimize the management data transferred to and fro in the network.

111

## 3. PERFORMANCE ANALYSIS

Many a times there arises a situation in network monitoring activities where one or two MIB variables representing some managed object may not be able to give good assessment of the issue (like performance degradation, signal failure, packet loss etc.). However, we need to access multiple parameters and then make an assessment. In order to carry out the performance analysis of the model presented in this work a bandwidth and throughput utilization function known as health function (HF) [6] which is essentially an aggregate of multiple MIB variable. In this five MIB-II [15] managed objects are combined to define the percentage $E(t)$ of IP packets discarded over the total number of packets sent within a specific time interval as shown in Equation 1

$$E(t) = \{(ipOutDiscards + ipOutNoRoutes + ipFragFails) / (ipOutRequests + ipForwDatagram)\}*100 \quad .....(1)$$

In simple client/server based SNMP model, either we have to issue five get requests to retrieve the values or at the best we would issue a single get request and the response would carry all the five parameters with their value. Back at manager level the value of $E(t)$ would be computed and necessary action would be planned. As suggested by Damianos Gavalas et al.[4] MAs can compute HFs locally thereby providing a way to semantically compress large amount of data (five variables, their IODs and return values and other SNMP Req/Res overheads) in a single value returned to the manager, thereby relieving it from processing NM data, while MAs state size remains as small as possible. But in case we have to compute this value for a collection of nodes, MAs would have to travel to these nodes and we need to bear the cost presented in [16].

The proposed mechanism offers an EMS database solution where we could issue a simple SQL query so that the needed values could be computed in a far less time and far less cost. For example, the $E(t)$ function shown in Equation (1), a simple SQL query similar to shown below can easily compute the function value at database level itself. With appropriate measures taken as schema design level like (indexes and data arrangement) similar SQL across series of nodes could be carried out at fraction of the cost that a mobile agent based flat bed model would incur.
Consider the HealthFunction given below as a view or some database schema table

| ipOutDiscard | ipOutNoRoutes | ipFragFails | ipOutRequest | ipForwDatagram |
|---|---|---|---|---|

The corresponding SQL query would be as shown below:

```
SELECT ipOutDiscards + ipOutNoRoutes + ipFragFails AS "Packet Discarded", ipOutRequests +    ipForwDatagram AS "Total Packets",Packet   Discarded / Total Packets AS "Health Function" From HealthFunction Where NE = `xxxx`;
```

## 4. NETWORK MANAGEMENT COST CALCULATION FOR THE PROPOSED MODEL

The Network management cost for IMASNM Model is discussed in details by A.K. Sharma et. al. [16] and has been used here as basis for cost calculation for the proposed model.  The network management cost calculation for the proposed model as shown in Equation (2) involves not only the management traffic cost i.e. (messages between the managers and sub-domain managers) but also the cost of setting up the managers as per the initial discovery of the network.

$$C_{TOTAL} = C_{SETUP} + C_{MGMTTR} \quad .....(2)$$

Where

$C_{SETUP}$: cost for discovering the network and deploying the managers – both M- SNLM & EMS.

$C_{MGMTTR}$: cost of a typical polling to know whole network status at top most level.

$C_{SETUP}$: the cost for discovering the network and deploying the managers – both M-SNLM & EMS is computed as shown in Equation (3)

$$C_{SETUP} = C_{MSNLM} + C_{EMS} \quad .....(3)$$

The cost of setting up the top level M-SNLMs managers as per the initial discovery of the network would be computed as shown in Equation (4)

$$C_{M\text{-}SNLM} = \sum_{h=1}^{L} \sum_{j=1}^{M-1} F\, h,j * MA\, Size \quad ......(4)$$

Where

*MASize:* Size of mobile manager,

*L:* Number of mother manager in the network,

*M:* Number of child managers in the sub domains of $h^{th}$ mother manager and

*Fh,j:* Sum of all the link cost coefficient between $h^{th}$ mother manager to $j^{th}$ child manager's sub domain.

The element managers would discover the domain element details in client/server mode and store the data in the EMS platform data for the agents visiting their platforms. The cost for setting up such platforms across the network would be computed as shown in Equation (5)

$$C_{EMS} = \{\sum_{i=1}^{N} K\, 0,1 * (Sreq + Sres)\} * M * L \quad .....(5)$$

Where

*N:* the no. of devices under the management of a given EMS,

*L:* Number of mother manager in the network,

*M:* Number of child managers in the sub domains of a mother manager

It may be noted that the deployment cost of the proposed model is higher than the IMASNM Model as cost of EMS gets added to the mobile-manager deployment. But as suggested in the proposed model architecture that it is a onetime cost and once the EMSs discover the network. They keep the data fresh through publish/subscribe interface. So for long term polling operations this cost could be ignored.

As discussed in IMASNM [16] model the management traffic cost for various polling operations is given as shown in Equation (6)

$$C_{MGMTTR} = \sum_{h=1}^{L} \sum_{j=0}^{M-1} Fh,j(MAres) + \sum_{j=1}^{Q} CQ \quad .....(6)$$

Where

*MAres:* Size of message sent by child to mother manager to report network health of its underlying domain,

*CQ:* Flat bed model cost for $Q^{th}$ domain's M-SNLM.

$$CQ = MDASize*(RQ+1)*KQ \quad .....(7)$$

Where

*MDASize:* Size of mobile data agent,

*RQ:* number of managed node is $Q^{th}$ domain and KQ : link coefficient of link's of $Q^{th}$ domain.

It may be noted that the flat bed cost increases in proportion to the cost coefficient of the links as well as number of nodes being managed in a given sub-domain. In the proposed model, the SQL performed by M-SNLMs would be local only and the cost incurred by flat bed model could be saved in total.

Hence, under the assumption that the cost of sql query would be far less than the cost of managing a domain by flat-bed model approach, the polling cost of the entire network could be computer as shown in Equation (8).

$$C_{\text{MGMTTR}} = \sum_{h=1}^{L} \sum_{j=0}^{M-1} Fh, j(MAres) \quad \quad \dots\dots(8)$$

Ignoring the one time setup cost, Equation (2) could be shown as

$C_{\text{TOTAL}} \approx C_{\text{MGMTTR}}$   (which is already optimised)

## 5. EXPERIMENT RESULTS

To support the management cost calculation presented in the previous section an experimental setup deploying Aglet version 2.0.2 as mobile agent platform, WebNMS Agent Toolkit Java Edition 6 for MIB design and manipulation, jdk1.6.0_13 as java runtime and MYSQL database was used. Experimental setup calculated the management cost in terms of time taken for management activity.

*For experiment results we consider three scenarios:*

*Scenario 1:* **Client/Server setup:** Figure 4 shows the simple client server model, where the static manager sends the SNMP request to agents, which sends the SNMP response back to manager. Node0 acts as M-SNLM for a sub-network and it collects information from nodes (node 1, node 2 …) in its domain using req/resp mechanism.

*Scenario 2:* **Flat-bed model:** Figure 6 shows the Flat-Bed model in which the mobile agents are responsible for accessing information by visiting one node to another. Node0 acting as M-SNLM dispatches the mobile agents for collecting the information from all the nodes in its domain and which they do in a flat-bed fashion and finally return back to Node0.

*Scenario 3:* **Hybrid model:** Figure 8 shows the "Hybrid model using EMS" in which the data is accesses from MYSQL database through SQL queries. The M-SNLM collects the information from nodes in its domain and keeps it in the database. The data is kept in sync with the nodes through simple publish/subscribe model. Whenever this M-SNLM is inquired from upper layer it simply queries the database and returns the result.

In the proposed experimental setup we calculate time for 50 get request & response for an Integer32 data type which is made by static manager (node 0) to static MIB agent (node 1, node 2, …) running on each node

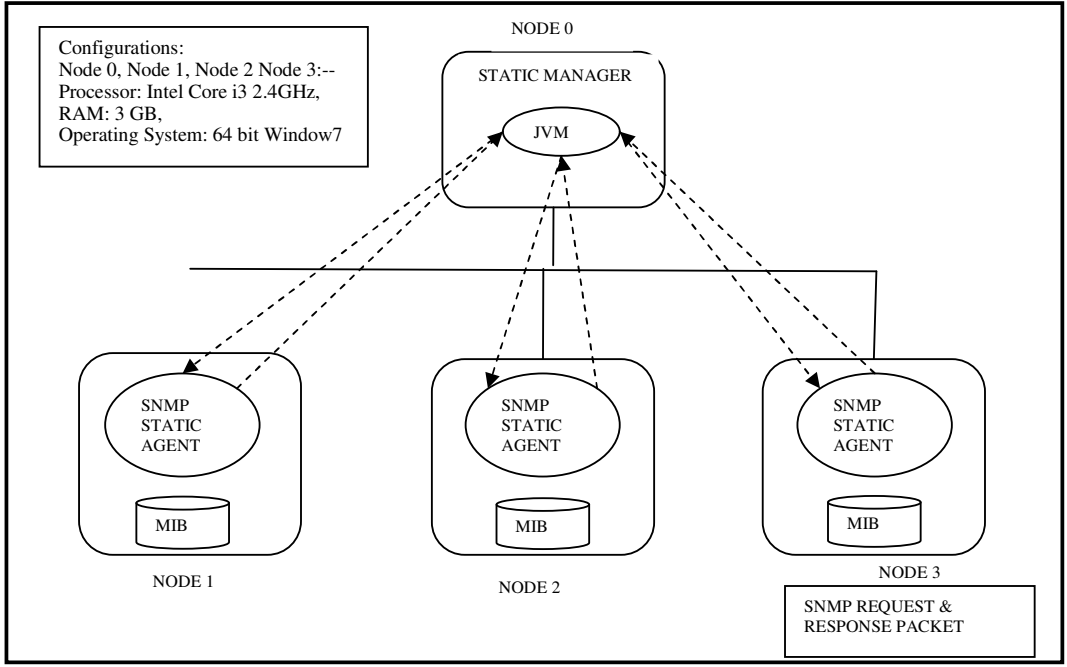

Figure 4.   Experimental Setup for C/S network management model

TABLE I.   EXPERIMENT RESULT OF CLIENT/SERVER BASED NETWORK MODEL

| S. no | Time1 | Time2 | Time3 |
|---|---|---|---|
| 1. | 179 | 249 | 113 |
| 2. | 159 | 145 | 122 |
| 3. | 159 | 268 | 150 |
| 4. | 239 | 131 | 210 |
| 5. | 141 | 129 | 112 |
| 6. | 422 | 144 | 88 |
| 7. | 166 | 117 | 120 |
| 8. | 71 | 195 | 322 |
| 9. | 185 | 188 | 169 |
| 10. | 91 | 137 | 227 |
| 11 | 152 | 164 | 100 |
| 12. | 398 | 198 | 279 |
| 13. | 221 | 188 | 138 |
| 14. | 199 | 109 | 130 |
| Average | 198 | 168 | 162 |
| Time1, Time2 and Time3 are respectively for Node1, Node2 and Node3. All times in milliseconds. | | | |

Figure 4 is experimental setup for scenario 1 which is a client/server based network management model. WebNMS toolkit is used for generating the client server applications for this setup. Table I show the experimental results of client server based model and figure 5 shows the corresponding graph.

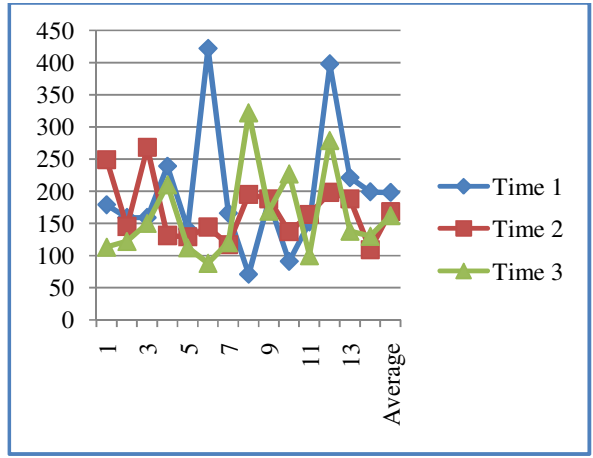

Figure 5. Graph of table I data values

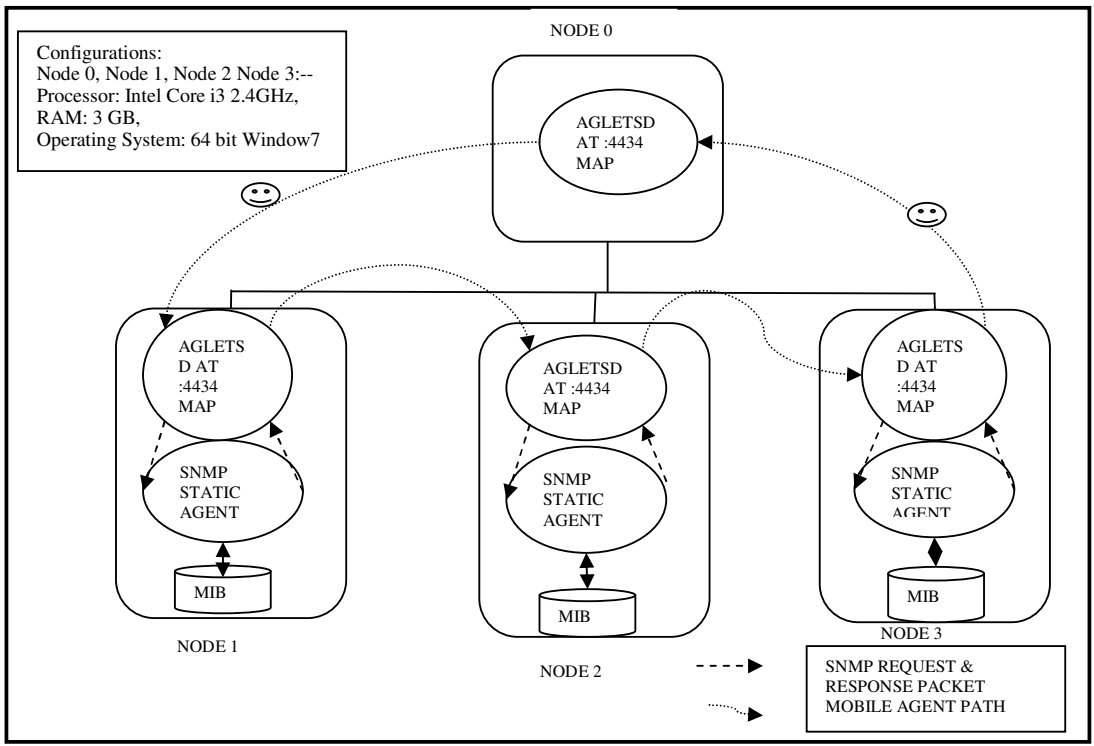

Figure 6. Experimental Setup for Flat- Bed model

Figure 6 is experimental setup for scenario 2 which is a MA based network management model. Aglet platform is used to send mobile agent from one node to another.

International Journal of Computer Networks & Communications (IJCNC) Vol.6, No.3, May 2014

TABLE II. Experiment result of Mobile agent based network model

| S. no | Time1 | Time2 | Time3 |
|---|---|---|---|
| 1. | 290 | 191 | 199 |
| 2. | 193 | 327 | 308 |
| 3. | 328 | 125 | 381 |
| 4. | 315 | 582 | 310 |
| 5. | 304 | 539 | 316 |
| 6. | 274 | 355 | 338 |
| 7. | 266 | 265 | 281 |
| 8. | 339 | 116 | 378 |
| 9. | 237 | 176 | 200 |
| 10. | 263 | 224 | 206 |
| 11 | 219 | 207 | 282 |
| 12. | 230 | 292 | 194 |
| 13. | 321 | 190 | 260 |
| 14. | 308 | 218 | 216 |
| 15. | 263 | 402 | 383 |
| Average | 276 | 280 | 283 |
| Time1, Time2 and Time3 are respectively for Node1, Node2 and Node3. All times in milliseconds. | | | |

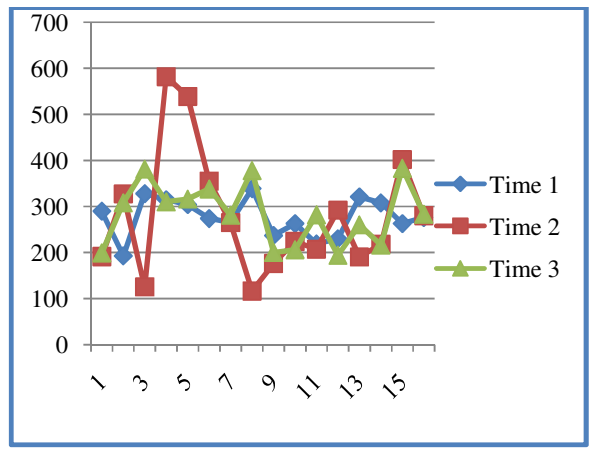

Figure 7. Graph of table II data values

Table II show the experimental results of mobile agent based model and figure 7 shows the corresponding graph.

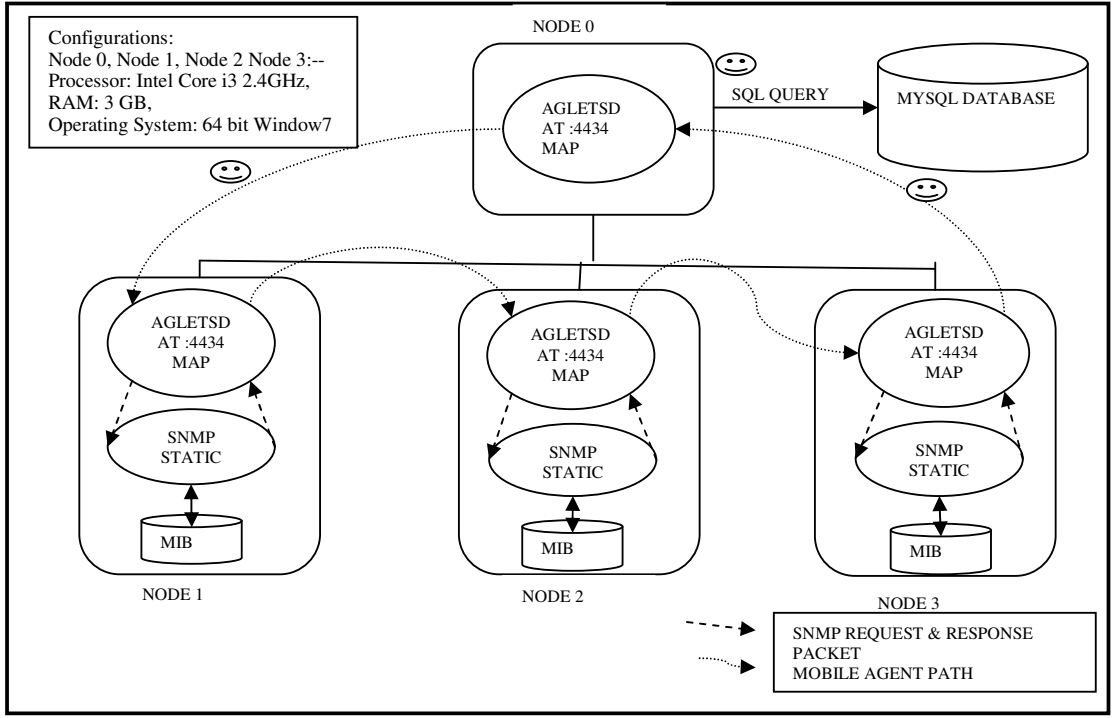

Figure 8. Experimental Setup for Hybrid model using ENMS

TABLE III. EXPERIMENT RESULT OF MOBILE AGENT BASED NETWORK MODEL ACCESSING MYSQL DATABASE

| S. no | Time1 | Time2 | Time3 |
|---|---|---|---|
| 1. | 122 | 39 | 31 |
| 2. | 47 | 31 | 31 |
| 3. | 143 | 46 | 47 |
| 4. | 31 | 25 | 19 |
| 5. | 16 | 31 | 15 |
| 6. | 30 | 23 | 18 |
| 7. | 27 | 19 | 20 |
| 8. | 44 | 33 | 32 |
| 9. | 32 | 31 | 31 |
| 10. | 16 | 31 | 16 |
| 11 | 44 | 35 | 34 |
| 12. | 35 | 31 | 16 |
| 13. | 30 | 20 | 20 |
| 14. | 30 | 27 | 20 |
| Average | 46 | 30 | 25 |
| Time1, Time2 and Time3 are respectively for Node1, Node2 and Node3. All times in milliseconds. | | | |

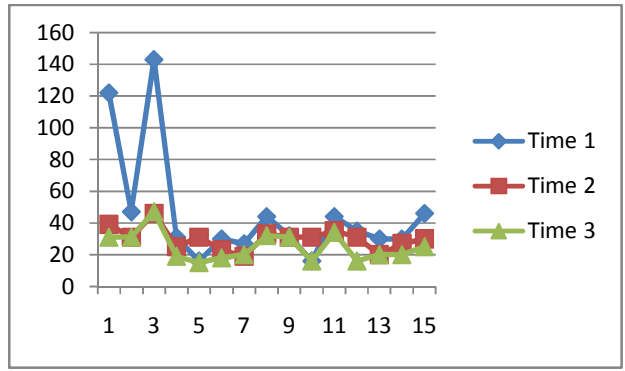

Figure 9. Graph of table 5.1 data values

In figure 8 node 0 is accessing values of 3 nodes i.e node1, node2, and node 3. Node 0 is also querying MYSQL database for the values which are stored in it, to do any operation. Table III shows time taken to access a value from MYSQL database.

## 6. CONCLUSIONS

The experimental results obtained in the previous section clearly show that client server based network management models takes less time as compare to mobile agent based models for a small size network. As the network grows in size client server based request and response put heavy load on network whereas mobile agents who operate locally at respective agent or node perform much better then client server. Moreover the model deploying database at the M-SNLM side performs even better than the mobile agent model. In client server model the management cost in terms of the data transferred to the Global manager for the whole network is directly proportional to the following factors:

- ➢ Number of requests and responses to fetch the data remotely
- ➢ Cost coefficient of the links on which information is exchanged.
- ➢ The number of MIB accessed.

Whereas in the proposed model, EMSs manage the domains locally thereby minimizing the cost incurred due to costly inter-domain link traversal of mobile agents. The cost of managing flat bed model is also taken away as SQL interface will fetch the needed data from EMSs database and it will be kept in sync with network elements by publish/subscribe interfaces and as the experimental results also proves that MYSQL queries takes less time as comapre to mobile agent. This simple experimental result shows that proposed EM based model would scale much better than Client/Server as well as IMASNM model.